\documentclass{revtex4}
\usepackage{amssymb}
\usepackage{amsfonts}
\usepackage{amsmath}
\usepackage{tikz,xcolor,hyperref}
\usepackage{nameref}

\begin{document}

\title{A note on "The Klein-Gordon oscillator in (1+2)-dimensions G\"{u}rses space-time backgrounds (Ann. Phys. (N. Y.) \textbf{404} (2019) 1)"}
\author{Omar Mustafa}
\email{omar.mustafa@emu.edu.tr}
\affiliation{Department of Physics, Eastern Mediterranean University, G. Magusa, north
Cyprus, Mersin 10 - Turkey.}
\author{Faizuddin Ahmed}
\email{faizuddinahmed15@gmail.com}
\affiliation{Department of Physics, University of Science and Technology Meghalaya, Ri-Bhoi, Meghalaya-793101, India.}

\vspace{0.5cm}

\begin{abstract}
\textbf{Abstract:}\ We revisit and discuss the KG-oscillator in the $(1+2)$-dimensional G\"{u}rses space-time studied in Ref. \cite{Re1} (F. Ahmed, Ann. Phys. (N. Y.) \textbf{404} (2019) 1). The modified oscillator frequency {\it i.e.,} $\tilde{\omega}^{2}=(\Omega^{2}\,E^{2}+\eta^{2})$ appeared in the eigenvalue equation is an energy-dependent parameter, and consequently, the results are accurately reported here. Moreover, we show some interesting spectroscopic features indulged within the very nature of G\"{u}rses space-time for the KG-G\"{u}rses oscillators.

\vspace{0.5cm}


\vspace{0.5cm} 


\end{abstract}

\maketitle

In a recent paper, one of us \cite{Re1} has investigated the Klein-Gordon (KG) particles in the $(1+2)$-dimensions G\"{u}rses space-time background \cite{Re2} described by the metric 
\begin{equation}
ds^2=-dt^2+dr^2-2\,\Omega\,r^2\,dt\,d\theta+r^2\,\left(1-\Omega^2\,r^2\right)\,d\theta^2=g_{\mu\nu}\,dx^{\mu}\,dx^{\nu};
\text{ }\mu
,\nu =0,1,2.  
\label{e1}
\end{equation}%
Under such space-time metric, the covariant and contravariant metric
tensors, respectively, read%
\begin{equation}
g_{\mu \nu }=\left( 
\begin{tabular}{ccc}
$-1\smallskip $ & $0$ & $-\Omega r^{2}$ \\ 
$0$ & $1\smallskip $ & $0$ \\ 
$-\Omega r^{2}$ & $\,0$ & $\,r^{2}\left( 1-\Omega ^{2}r^{2}\right) $%
\end{tabular}%
\right) \Longleftrightarrow g^{\mu \nu }=\left( 
\begin{tabular}{ccc}
$\left( \Omega ^{2}r^{2}-1\right) $ & $0\smallskip $ & $-\Omega $ \\ 
$0$ & $1\smallskip $ & $0$ \\ 
$-\Omega $ & $\,0$ & $\,1/r^{2}$%
\end{tabular}%
\right) \text{ };\text{ \ }g=\det \left( g_{\mu \nu }\right) =-r^{2}.
\label{e4}
\end{equation}%
Then the corresponding KG-equation is given by%
\begin{equation}
\frac{1}{\sqrt{-g}}\partial _{\mu }\left( \sqrt{-g}g^{\mu \nu }\partial
_{\nu }\Psi \right) =m^{2}\Psi .  \label{e5}
\end{equation}%
At this point, the KG-oscillator is introduced through Moshinsky and Szczepaniak recipe \cite{Re3} so that the momentum operator takes its non-minimal coupling form
\begin{equation}
p_{\mu }\longrightarrow p_{\mu }+i\mathcal{F}_{\mu }\Longrightarrow \partial_{\mu }\longrightarrow \partial _{\mu }+\mathcal{F}_{\mu },  
\label{e6}
\end{equation}%
where $\mathcal{F}_{\mu }=\left( 0,\mathcal{F}_{r},0\right) =\left(0, m\,\omega\,r, 0\right) $ as proposed in \cite{Re3,Re4,Re5,Re6,Re8}. We shall, however, define $\mathcal{F}_{\mu }=\left( 0,\eta \,r,0\right)$ to avoid the confusion between $m^{2}\equiv \left( mc^{2}\right) ^{2}$ in (\ref{e5}) and $m$ in $\mathcal{F}_{\mu }=\left( 0,m\,\omega r,0\right) $ which identifies the non-relativistic mass of the quantum particle (as in Schr\"{o}dinger equation). To retrieve Ahmed's results, we substitute $\eta =m\,\omega$, therefore. Under such settings, one rewrites (\ref{e5}) as 
\begin{equation}
\frac{1}{\sqrt{-g}}\left( \partial _{\mu }+\mathcal{F}_{\mu }\right) \left[\sqrt{-g}\,g^{\mu\nu }\left( \partial_{\nu }-\mathcal{F}_{\nu }\right) \Psi\right]=m^{2}\Psi,
\label{e7}
\end{equation}%
to obtain%
\begin{equation}
\left\{ -\partial _{t}^{2}+\left( \Omega \,r\,\partial _{t}-\frac{1}{r}%
\partial _{\theta }\right) ^{2}+\partial _{r}^{2}+\frac{1}{r}\partial
_{r}-\eta ^{2}r^{2}-2\eta -m^{2}\right\} \Psi =0.  \label{e8}
\end{equation}%
Then the substitution of $\Psi =\Psi \left( t,r,\theta \right) =\exp \left(-i\,E\,t+i\,\ell\,\theta \right)\,\psi\left( r\right) $ (where $\ell $ denotes the magnetic quantum number $\ell =0,\pm 1,\pm 2,\cdots $) would result in%
\begin{equation}
\psi ^{^{\prime \prime }}\left( r\right) +\frac{1}{r}\psi ^{^{\prime}}\left( r\right) +\left( \lambda -\tilde{\omega}^{2}r^{2}-\frac{\ell ^{2}}{r^{2}}\right) \psi \left( r\right) =0,  
\label{e09}
\end{equation}%
where 
\begin{equation}
\lambda =E^{2}-2\,\Omega\,\ell\,E-m^{2}-2\,\eta,\text{ }\tilde{\omega}=\sqrt{\Omega^{2}\,E^{2}+\eta^{2}}.
\label{e010}
\end{equation}%
We observe that equation (\ref{e09}) is in the form of the two-dimensional radial Schr\"{o}dinger oscillator and admits the textbook solution%
\begin{equation}
\psi \left( r\right) \sim r^{\left\vert \ell \right\vert }\exp \left( -\frac{%
\left\vert \tilde{\omega}\right\vert r^{2}}{2}\right) L_{n_{r}}^{\left\vert
\ell \right\vert }\left( \left\vert \tilde{\omega}\right\vert r^{2}\right) 
\label{e011}
\end{equation}%
as the eigenfunctions, and 
\begin{equation}
\lambda =2|\tilde{\omega}|\left( 2n_{r}+|\ell |+1\right)   \label{e012}
\end{equation}%
as the eigenvalues, where $n_{r}=0,1,2,\cdots $ is the radial quantum number. Consequently, using (\ref{e010}) and (\ref{e012}), one would obtain the following eigenvalue polynomial equation %
\begin{equation}
E^{2}_{n_r,\ell}-2\,\Omega\,\ell\,E_{n_r,\ell}-C_{n_r,\ell}^{2}=0,  \label{e013}
\end{equation}%
where%
\begin{equation}
\text{ \ }C_{n_{r},\ell }^{2}=2|\tilde{\omega}|\,\left( 2n_{r}+|\ell|+1\right) +m^{2}+2\,\eta 
\label{e014}
\end{equation}%
which is exactly the same expression as equation (20), along with (22), reported in Ref. \cite{Re1} using the parametric Nikiforov-Uvarov method.

However, the following discussion was missing in Ref. \cite{Re1} which we feel necessary to explain. The modified oscillator frequency $\tilde{\omega}$, given by (\ref{e010}) with $|\tilde{\omega}|=|\Omega\,E|\sqrt{1+\eta^{2}/\Omega ^{2}E^{2}}$, is actually an energy-dependent parameter, and $C_{n_{r},\ell }$'s of (\ref{e014}) are given by
\begin{equation}
C_{n_{r},\ell }^{2}=C_{n_{r},\ell }^{2}\left( E_{n_{r},\ell }\right) =2\,\sqrt{\Omega^2\,E^{2}_{n_r,\ell}+\eta^2}\left(2\,n_r+|\ell |+1\right)+m^{2}+2\,\eta .  
\label{e015}
\end{equation}%
Therefore, for any $n_{r}\geq 0$ the results starting from equations (25) to (30) in Ref. \cite{Re1} seem insecure (since they are in an incomplete form). For example, $E_{0,\ell }$ given in equation (25) in the original work is explicitly given by
\begin{equation}
E_{0,\ell }=\Omega\,\ell +\sqrt{\Omega^{2}\,\ell^{2}+2\,\sqrt{\Omega^{2}\,E_{0,\ell}^{2}+\eta^{2}}\left( |\ell |+1\right) +m^{2}+2\,\eta }
\label{e016}
\end{equation}%
for $n_r=0$ and so on. Actually, the eigenvalue equation (\ref{e013}) is a polynomial of $E_{n_r,\ell}$ given by
\begin{eqnarray}
    &&E^{2}_{n_r,\ell}-2\,\Omega\,\ell\,E_{n_r,\ell}-m^2-2\,\eta=2\,\sqrt{\Omega^{2}\,E^{2}_{n_r,\ell}+\eta^{2}}\,\left(2\,n_{r}+|\ell|+1\right)\nonumber\\ \Rightarrow
    && \left(E^{2}_{n_r,\ell}-2\,\Omega\,\ell\,E_{n_r,\ell}-m^2-2\,\eta\right)^2 =4\,\left(\Omega^{2}\,E^{2}_{n_r,\ell}+\eta^{2}\right)\,\left(2\,n_{r}+|\ell|+1\right)^2\nonumber\\ \Rightarrow
    &&f_4\,E^{4}_{n_r,\ell}+f_3\,E^{3}_{n_r,\ell}+f_2\,E^{2}_{n_r,\ell}+f_1\,E_{n_r,\ell}+f_0=0
    \label{ss}
\end{eqnarray}
where $f_i\quad (i=0 \ldots 4)$ are non-zero parameter. The real solution of the above polynomial (\ref{ss}) gives us the exact energy eigenvalue expression $E_{n_r, \ell}$ of the KG-oscillator field associated with the modes $\left\{n_r, \ell\right\}$ in the G\"{u}rses space-time. 

Nevertheless, it would be interesting to mention that the G\"{u}rses space-time metric (\ref{e1}) provides the KG-oscillators (hence KG-G\"{u}rses oscillators) as a natural G\"{u}rses space-time byproduct. Namely, setting $\eta =0$ (i.e., retrieving the usual momentum operator in (\ref{e6})) would result that $\tilde{\omega}^{2}=\Omega ^{2}E^{2}$ in (\ref{e09}) and the corresponding quadratic energy form of equation (\ref{e013}) now reads%
\begin{equation}
E^{2}-2\,\Omega\,\ell\,E-2\,|\Omega\,E|\,\left(2\,n_{r}+|\ell|+1\right)-m^{2}=0.
\label{e017}
\end{equation}%
This energy relation is by itself interesting in the sense that $E$ denotes the energies for both KG-particles and anti-particles, {\it i. e.,} $E=E_{+}=+|E|$ for particles and $E=E_{-}=-|E|$ for anti-particles. Moreover, the vorticity parameter $\Omega $ denotes rotation of the G\"{u}rses universe so that $\Omega =\pm|\Omega |$ to imply that $\Omega _{+}=|\Omega |$ and $\Omega _{-}=-|\Omega |$. Under such assumptions, one would obtain%
\begin{equation}
E_{\pm }^{2}-2\,\Omega _{\pm }E_{\pm }\,\tilde{n}_{+}-m^{2}=0;\;\tilde{n}%
_{+}=2n_{r}+\left\vert \ell \right\vert +\ell \,+1,  \label{e018}
\end{equation}%
for $\left\vert \Omega E\right\vert =\Omega _{\pm }E_{\pm }$, and 
\begin{equation}
E_{\pm }^{2}+2\,\Omega _{\mp }E_{\pm }\,\tilde{n}_{-}\,-m^{2}=0;\;\tilde{n}%
_{-}=2n_{r}+\left\vert \ell \right\vert -\ell \,+1.  \label{e019}
\end{equation}%
for $\left\vert \Omega E\right\vert =-\Omega _{\mp }E_{\pm }$. Which would
allow us to cast%
\begin{equation}
E_{\pm }=\Omega _{\pm }\,\tilde{n}_{+}\pm \sqrt{\Omega ^{2}\tilde{n}%
_{+}^{2}+m^{2}}\Rightarrow \left\{ 
\begin{tabular}{l}
$E_{+}=\Omega _{\pm }\,\tilde{n}_{+}+\sqrt{\Omega ^{2}\tilde{n}_{+}^{2}+m^{2}%
}$ \\ 
$E_{-}=\Omega _{\pm }\,\tilde{n}_{+}-\sqrt{\Omega ^{2}\tilde{n}_{+}^{2}+m^{2}%
}$%
\end{tabular}%
\right. ,  \label{e020}
\end{equation}%
for $\left\vert \Omega E\right\vert =\Omega _{\pm }E_{\pm }$ and%
\begin{equation}
E_{\pm }=-\Omega _{\mp \,}\tilde{n}_{-}\,\pm \sqrt{\Omega ^{2}\tilde{n}%
_{-}^{2}+m^{2}}\Rightarrow \left\{ 
\begin{tabular}{l}
$E_{+}=-\Omega _{-\,}\tilde{n}_{-}\,+\sqrt{\Omega ^{2}\tilde{n}_{-}^{2}+m^{2}%
}$ \\ 
$E_{-}=-\Omega _{+\,}\tilde{n}_{-}\,-\sqrt{\Omega ^{2}\tilde{n}_{-}^{2}+m^{2}%
}$%
\end{tabular}%
\right. .  \label{e021}
\end{equation}%
However, it is more interesting to express the energies in terms of $\Omega
_{+}$ and $\Omega _{-}$so that%
\begin{equation}
E_{\pm }^{\left( \Omega _{+}\right) }=\pm \left\vert \Omega \right\vert \,%
\tilde{n}_{\pm }\pm \sqrt{\Omega ^{2}\tilde{n}_{\pm }^{2}+m^{2}},
\label{e022}
\end{equation}%
for positive vorticity parameter, and%
\begin{equation}
E_{\pm }^{\left( \Omega _{-}\right) }=\pm \left\vert \Omega \right\vert \,%
\tilde{n}_{\mp }\pm \sqrt{\Omega ^{2}\tilde{n}_{\mp }^{2}+m^{2}}.
\label{e023}
\end{equation}%
for negative vorticity parameter. Notably, we observe that $\tilde{n}_{\pm }\left(
\ell =\pm \left\vert \ell \right\vert \right) =\tilde{n}_{\mp }\left( \ell
=\mp \left\vert \ell \right\vert \right) $ which would in effect introduce
the so called vorticity-energy correlations so that $E_{\pm }^{\left( \Omega
_{+}\right) }\left( \ell =\pm \left\vert \ell \right\vert \right) =E_{\pm
}^{\left( \Omega _{-}\right) }\left( \ell =\mp \left\vert \ell \right\vert
\right) $.

On the other hand, the inclusion of the coupling $\xi $ of the gravitational
field with the background curvature $R=2\Omega ^{2}$ would imply that%
\begin{equation}
E_{\pm }^{\left( \Omega _{+}\right) }=\pm \left\vert \Omega \right\vert \,%
\tilde{n}_{\pm }\pm \sqrt{\Omega ^{2}\tilde{n}_{\pm }^{2}+m^{2}+2\xi \Omega
^{2}},  \label{e024}
\end{equation}%
for the energies with positive vorticity in (\ref{e022}), and 
\begin{equation}
E_{\pm }^{\left( \Omega _{-}\right) }=\pm \left\vert \Omega \right\vert \,%
\tilde{n}_{\mp }\pm \sqrt{\Omega ^{2}\tilde{n}_{\mp }^{2}+m^{2}+2\xi \Omega
^{2}}.  \label{e025}
\end{equation}%
for the energies with negative vorticity in (\ref{e023}). We again observe
the vorticity-energy correlations $E_{\pm }^{\left( \Omega _{+}\right)
}\left( \ell =\pm \left\vert \ell \right\vert \right) =E_{\pm }^{\left(
\Omega _{-}\right) }\left( \ell =\mp \left\vert \ell \right\vert \right) $
for the energies in (\ref{e024}) and (\ref{e025}).

Finally, the current note is intended to show some interesting spectroscopic features indulged within the G\"{u}rses spacetime structure, especially for the KG-G\"{u}rses equation. It is worthwhile mentioning that in the original work \cite{Re1}, the vorticity parameter $\Omega$ is chosen to be positive and forgot to interpreted the modified oscillator frequency $\tilde{\omega}$ for which the exact energy is now possible to obtain from the polynomial equation (\ref{ss}) (but not the equation (21) in \cite{Re1} written in an incomplete form). In this note, we have shown that since the modified oscillator frequency $\tilde{\omega}$ given by (\ref{e010}) is an energy-dependent parameter, and hence, the exact energy expression of the KG-oscillator can be obtained by solving the polynomial equation (\ref{ss}) of the real solution. Consequently, the results reported in the original paper \cite{Re1} for different modes, such as $n_r=0,1,2$ are insecure and incomplete.

\vspace{0.5cm}

\textbf{Data Availability Statement:} No new data generated or analyzed in this study.

\vspace{0.5cm}

\textbf{Conflicts of Interest:} There is no conflict of interests.

\end{document}